\documentclass[a4paper,11pt]{article}
\usepackage{pos}

\newcommand{\Tr}[1]{\text{Tr}\left[#1\right]}
\newcommand{\Z}{{\mathbb Z}}
\newcommand{\realpart}{\text{Re}}

\title{Effective $\Z_{3}$ model for finite-density QCD with tensor networks}
\author[a]{Jacques Bloch}
\author[a,b]{Robert Lohmayer}
\author[a]{Sophia Schwei{\ss}}
\author*[c]{Judah Unmuth-Yockey}

\affiliation[a]{Institute for Theoretical Physics, University of Regensburg, \\ Regensburg, Germany}

\affiliation[b]{RCI Regensburg Center for Interventional Immunology, \\ Regensburg, Germany}

\affiliation[c]{Department of Theoretical Physics, Fermi National Accelerator Laboratory, \\ Batavia, IL}

\emailAdd{jacques.bloch@ur.de}
\emailAdd{robert.lohmayer@ur.de}
\emailAdd{Sophia.Schweiss@stud.uni-regensburg.de}
\emailAdd{jfunmuthyockey@gmail.com}

\abstract{The tensor renormalization group is a promising numerical method used to study lattice statistical field theories. However, this approach is computationally expensive in 2+1 and 3+1 dimensions. Here we use tensor renormalization group methods to study an effective three-dimensional $\Z_{3}$ model for the heavy-quark, high-temperature, strong-coupling limit of single-flavor 3+1 dimensional quantum chromodynamics. Our results are cross-checked using the worm Monte Carlo algorithm. We present the phase diagram of the model through the measurement
of the Polyakov loop, the nearest-neighbor Polyakov loop correlator, and their susceptibilities. The tensor renormalization group results are in good agreement with the literature.}

\FullConference{%
 The 38th International Symposium on Lattice Field Theory, LATTICE2021
  26th-30th July, 2021
  Zoom/Gather@Massachusetts Institute of Technology
}

\begin{document}
\maketitle

\section{Introduction}
Tensor-network methods provide a tool to approximately coarse-grain a classical partition function.  Moreover, in certain cases, the method is immune to the sign problem, since there is no probabilistic sampling involved.  Here we apply two tensor methods---the higher-order tensor renormalization group (HOTRG)~\cite{hotrg}, and the the triad tensor renormalization group (TTRG)~\cite{triad}---in conjunction with the worm Monte Carlo algorithm~\cite{worm}, to study an effective theory for one-flavor, finite-density, 3+1-dimensional quantum chromodynamics (QCD).  The effective theory has a sign problem in the original formulation, however this can be overcome with a change of variables, making the model ideal to test the tensor methods against the Monte Carlo results.  We first derive the effective theory, then provide a tensor representation of the theory, and finally report our results.

\section{Single-flavor, finite-density QCD}
    The starting action is the standard lattice Wilson gauge action, and Wilson single-flavor fermion action.
    The action has two parts, $S = S_{g} + S_{f}$, with the {gauge action} given by
    \begin{align}
    S_g  &= -\frac{\beta}{3}\sum_{x=1}^{N} \left[ \frac{a}{a_t} \sum_{i=1}^{3} \realpart [\Tr{U_{x,i4}}]
        + \frac{a_t}{a} \sum_{i < j=1}^{3} \realpart[ \Tr{U_{x, i j}}] \right]
    \label{GaugeAction}
    \end{align}
    and the {fermion action} by
    \begin{align}
    S_{f} = \sum_{x,y} \bar{\psi}_{x} M_{xy} \psi_{y}, \quad M &= \mathbf{1} - \kappa\frac{a_t}{a} H_{s} - \kappa H_{t}
    \end{align}
    with the spatial and temporal hoppings
    \begin{align}
    H_{s} = \sum_{i=1}^3 \left(T_i^+ + T_i^-\right) ,\quad
    H_{t} = e^{\tilde{\mu} a_t} T_4^+ + e^{-\tilde{\mu} a_t} T_4^-
    \label{Ht}
    \end{align}
    and
    \begin{align}
    (T_\nu^{\pm})_{xy} &= (1\pm\gamma_{\nu}) U_{x,\pm \nu}\delta_{y,x\pm\hat{\nu}} .
    \end{align}
    Here, the gauge coupling enters through $\beta = 6/g^2$ and we identify $\kappa$ as the fermion hopping parameter. The chemical potential is $\tilde{\mu}$, and $a$ and $a_t$ are the spatial and temporal lattice spacings, respectively.  In anticipation of deriving the effective theory, we define the temperature as given by the inverse  physical extent of the lattice,
    \begin{equation}
        T = \frac{1}{N_{t} a_{t}},
    \end{equation}
    with $N_{t}$ the number of lattice sites in the temporal direction.  We work on a finite lattice with anti-periodic boundary conditions in the temporal direction, and periodic boundary conditions in the spatial directions.
    
\section{The effective action}
    The effective action we study corresponds to a particular limiting case.  To derive it we follow Ref.~\cite{degrand} closely.  We consider the strong-coupling, high-temperature limit, where we take the temporal extent of the lattice very short---in this case we set $N_{t} = 1$---and shrink the temporal lattice spacing.  This corresponds to
     \begin{align}
        \beta \ll 1, \quad a_{t} \ll 1.
    \end{align}
    In addition we take the large-chemical-potential, large-quark-mass limit, which in the action corresponds to
    \begin{align}
        \tilde{\mu} \gg 1, \quad \kappa \ll 1.
    \end{align}
    In $S_{g}$ spatial plaquettes are suppressed by a factor of $a_{t}^{2}$ relative to the temporal plaquettes, and in $S_{f}$ spatial hopping is suppressed.  Since the confinement-deconfinement transition is believed to be related to center-symmetry breaking, we replace the $SU(3)$ group elements by elements of the center of the group, $\Z_{3}$, which is an Abelian group.  For $S_{g}$ we find
    \begin{align}
        S_{g} \rightarrow -\frac{\beta}{2 a_{t}} \sum_{x}\sum_{\nu=1}^{3} P^{*}_{x} P_{x+\hat{\nu}} +
        \text{c.c.}
    \end{align}
    and for $S_{f}$
    \begin{align}
        S_{f} \rightarrow -\kappa \sum_{x}\left[  e^{\mu} P_{x} + e^{-\mu} P^{*}_{x} \right]
    \end{align}
    with $P_{x} \in \Z_{3}$, which can be interpreted as a {Polyakov loop} in the short, compactified temporal direction, and $\mu = a_{t} \tilde{\mu}$. This results in an effective action
    % \begin{equation}
    %     \boxed{S_{\text{eff}} = -\sum_{x} \left[ \tau \sum_{\nu=1}^{3} \left( P^{*}_{x} P_{x+\hat{\nu}} +
    %     \text{c.c.} \right) + \left( \eta P_{x} + \bar{\eta} P^{*}_{x} \right) \right]}
    % \end{equation}
    \begin{equation}
        S_{\text{eff}} = -\sum_{x} \left[ \tau \sum_{\nu=1}^{3} \left( P^{*}_{x} P_{x+\hat{\nu}} +
        \text{c.c.} \right) + \left( \eta P_{x} + \bar{\eta} P^{*}_{x} \right) \right]
    \end{equation}
    with $\tau \equiv \beta / 2 a_{t}$ and $\eta \equiv \kappa e^{\mu}$ and $\bar{\eta} = \kappa e^{-\mu}$.  A graphical illustration of the Polyakov loop variables on the lattice can be seen on the left-hand side of Fig.~\ref{fig:pictures}.  The effective theory lives in three Euclidean dimensions, and is to be interpreted as a statistical theory, in equilibrium, in three spatial dimensions.

%     \column{0.33}
    \section{The tensor network}
    \label{sec:tensornetwork}
    
    \begin{figure}
        \includegraphics[width=0.49\textwidth]{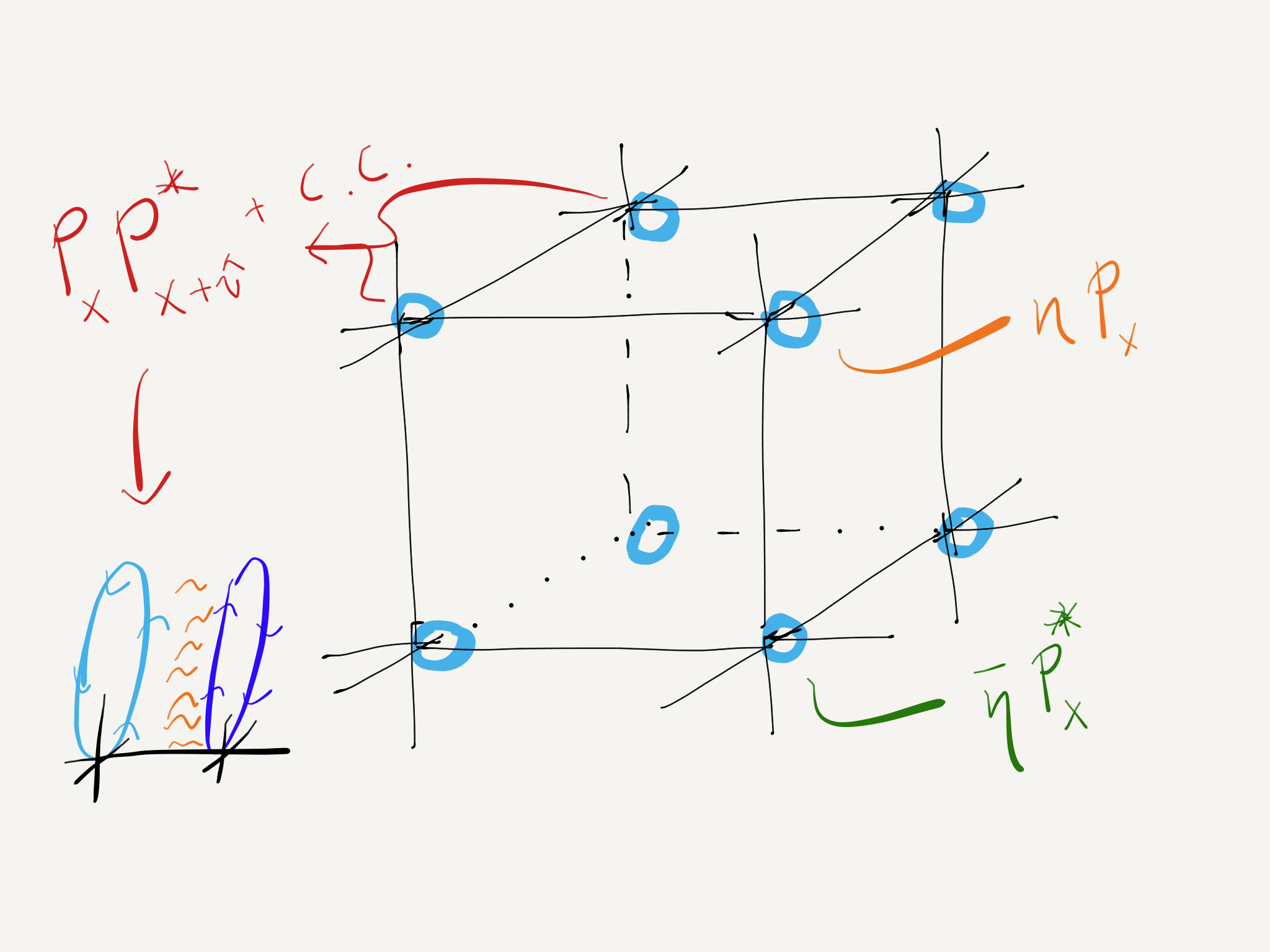}
    %     \end{figure}
    % \begin{figure}
        % \hfill
        \includegraphics[width=0.49\textwidth]{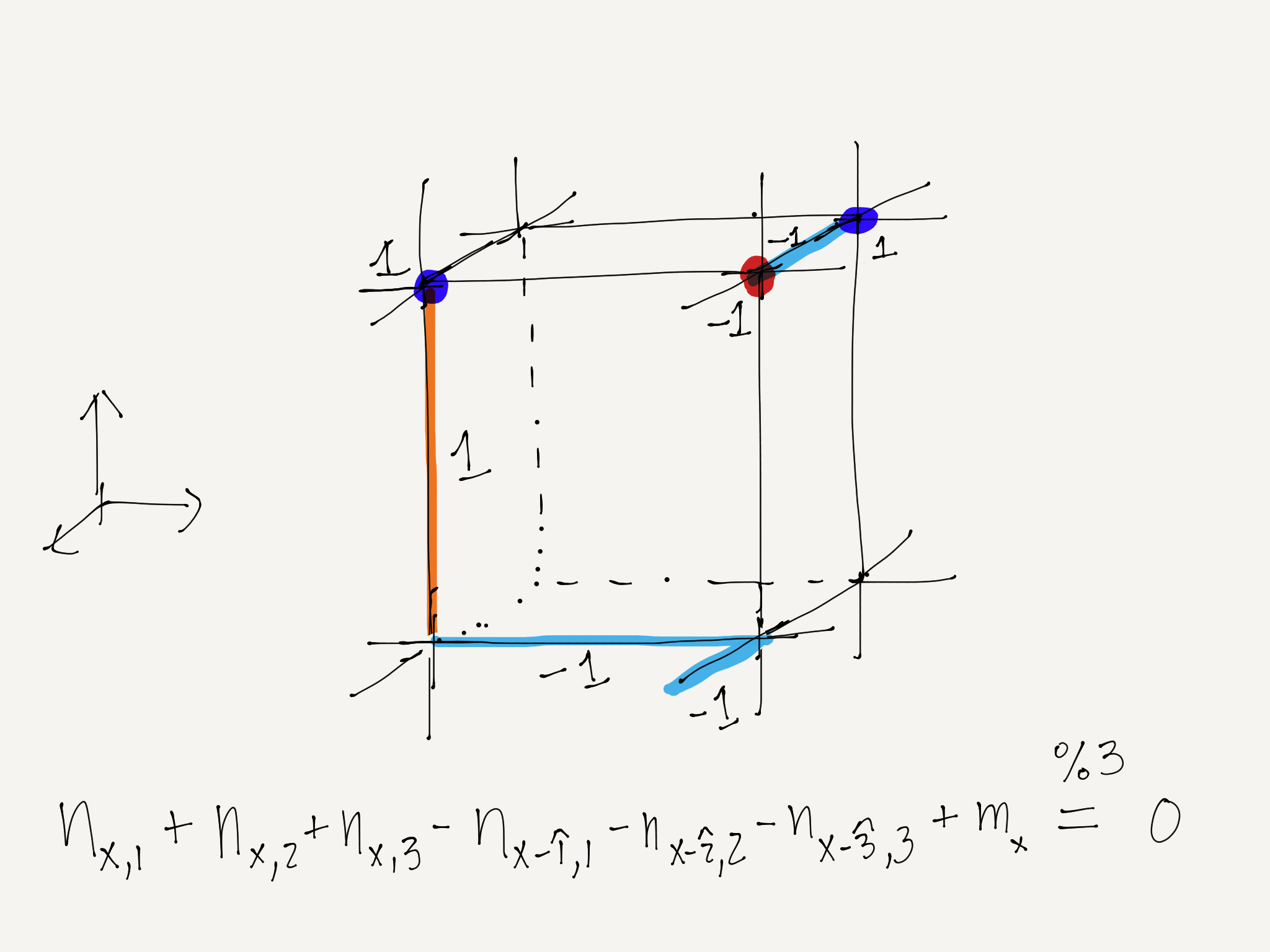}
        \caption{(left) An illustration of the original degrees of freedom.  At each site there is a $\Z_{3}$ element (a Polyakov loop) which interacts with its nearest neighbors.  (right) In the Fourier basis, there are ``charges'' and ``currents'' on the sites, and links, respectively, which must obey a conservation-law constraint at each site.}
        \label{fig:pictures}
    \end{figure}
    The partition function for the effective action describing
    this theory of interacting $\Z_{3}$ ``spins'' is given by,
    \begin{align}
    Z = \sum_{\{P\}} e^{-S_{\text{eff}}},
    \end{align}
    where the sum, $\sum_{\{P\}}$, means a sum over all possible configurations of the $\Z_{3}$ variables. This can be recast to a theory of ``currents'' and ``charges'' by Fourier expansion of the elementary Boltzmann weights~\cite{gattringer},
    \begin{align}
        \exp(\tau[P_x P_{x+\hat\nu}^* + \text{c.c.}]) &= C \hspace{-3ex} \sum_{n_{x,\nu}\in\{-1,0,1\}} \hspace{-3ex}
B^{|n_{x,\nu}|} (P_x P_{x+\hat\nu}^*)^{n_{x,\nu}}
    \end{align}
     and
    \begin{align}
        \exp(\eta P_x + \bar\eta P_x^*) &= \hspace{-2ex} \sum_{m_x\in\{-1,0,1\}} \hspace{-2ex} V_{m_x} P_x^{m_x}
    \end{align}
    where $B$ and $V$ can be found by inverting the above.  These expansions allow the original $\Z_{3}$ variables to be summed over, leaving only the $n$ and $m$ integer variables as the new degrees of freedom.  The resulting partition function is written as,
    \begin{align}
        Z = \sum_{\{n\}} \sum_{\{m\}}
        \left( \prod_{x,\nu} B_{n_{x,\nu}}(\tau) \right)
        \left( \prod_{x} V_{m_{x}}(\eta, \bar{\eta}) \Delta^{(3)}(j_{x}) \right),
    \end{align}
    with $j_{x} = \sum_{\nu=1}^{3} (n_{x,\nu} - n_{x-\hat{\nu},\nu}) + m_{x}$, and $\Delta^{(3)}(j_{x}) = \delta_{j_{x}\text{mod}3,0}$.  A graphical illustration of the new degrees of freedom on the lattice can be seen in Fig.~\ref{fig:pictures} on the right-hand side.

    A {tensor network} can be built from the tensor,
    \begin{align}
    % \nonumber
    \label{eq:tensor}
    T^{(x)}_{n_{x-\hat{1},1} n_{x,1} n_{x-\hat{2},2} n_{x,2} n_{x-\hat{3},3} n_{x,3}} = \sqrt{B_{n_{x-\hat{1},1}}B_{n_{x,1}}
        B_{n_{x-\hat{2},2}}
        B_{n_{x,2}}
        B_{n_{x-\hat{3},3}}B_{n_{x,3}}(\tau)} 
        V_{m_x}(\eta, \bar{\eta}) \Delta^{(3)}(j_{x})
        % T_{n_{x,1} n_{x,2} n_{x,3} n_{x-\hat{1},1} n_{x-\hat{2},2} n_{x-\hat{3},3}}
        % = \left[ B_{n_{x,1}} B_{n_{x,2}} B_{n_{x,3}} B_{n_{x,\nu}}
        % B_{n_{x,\nu}} B_{n_{x,\nu}} \right]^{\frac{1}{2}}
    \end{align}
    and the partition function written as a tensor trace
    \begin{align}
        Z = \Tr{\prod_{x} T^{(x)}}.
    \end{align}
    We calculate the average {Polyakov loop} and average {nearest-neighbor correlator}, and their susceptibilities,
    \begin{align}
        \langle P \rangle = \frac{1}{V} 
        \frac{\partial \ln Z}{\partial \eta}, \quad \langle E \rangle =
        - \frac{1}{V} \frac{\partial \ln Z}{\partial \tau}
    \end{align}
    \begin{align}
        \chi_{P} = V (\langle P^{2} \rangle - \langle P \rangle^{2}),
        \quad
        \chi_{E} = V (\langle E^{2} \rangle - \langle E \rangle^{2}).
    \end{align}
    We do this using ``impure'' tensors~\cite{imp-gu,MORITA201965}, and finite differences of the logarithm of the partition function.
    % The partition function can be written entirely interms of the local tensor above,
    % \begin{align}
    %     Z = \Tr{\prod_{x} T^{(x)}}.
    % \end{align}

    \section{Numerical methods}
    
    \begin{figure}
    \centering
        \includegraphics[trim={0 2cm 0 2cm},width=0.4\textwidth]{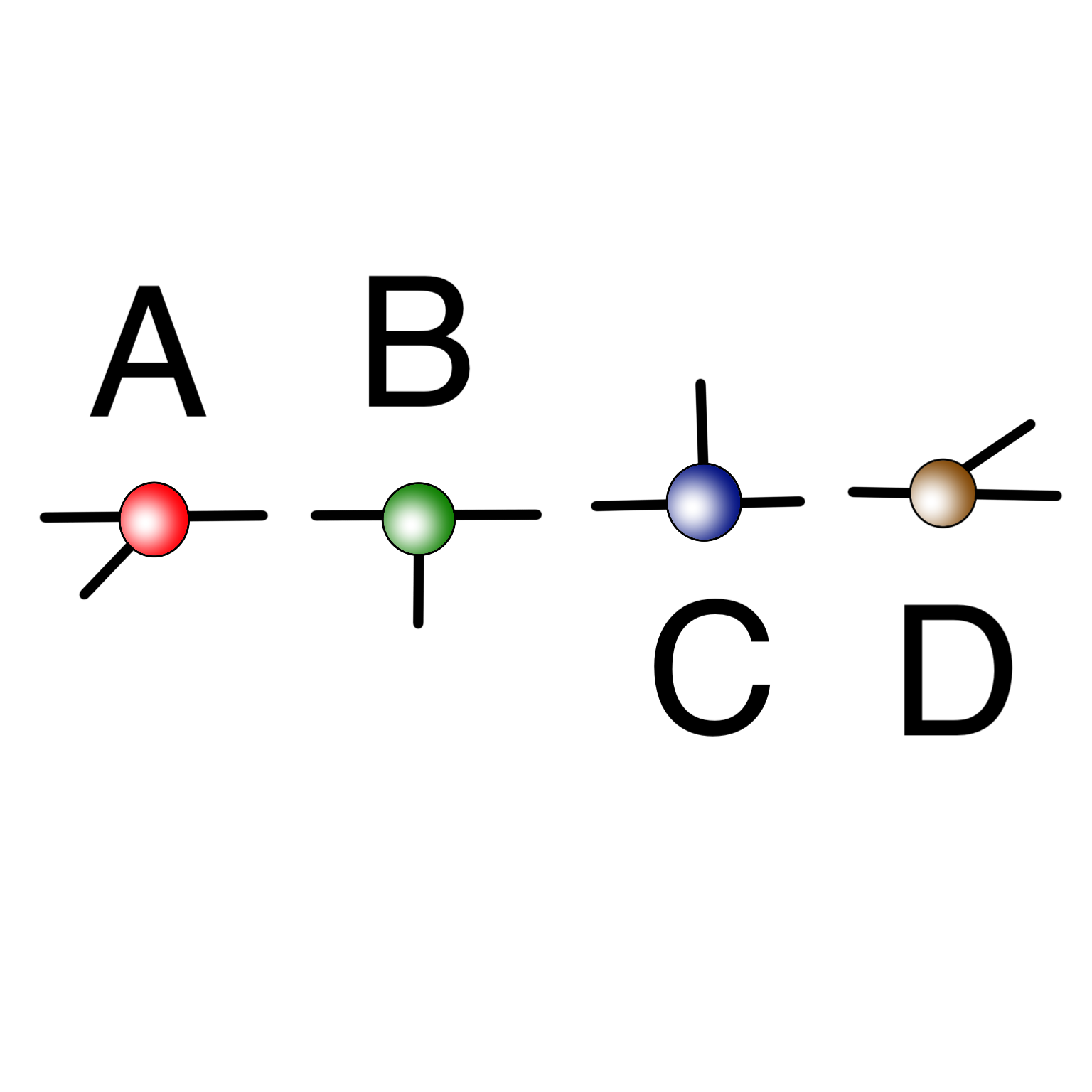}
        \caption{A graphical illustration of the decomposition of the fundamental six-indexed tensor into four triad tensors.}
        \label{fig:namedtriads}
    \end{figure}
    Using $T$ from Eq.~\eqref{eq:tensor} we contract the tensor network using the HOTRG.  This can be done in the original HOTRG prescription~\cite{hotrg}, or with additional approximations using the TTRG~\cite{triad}.  Within the TTRG,
    The $T$ tensor can be decomposed into four, three-indexed tensors called triads,
    \begin{align}
        T_{ijklmn} = \sum_{\alpha,\beta,\gamma}
        A_{ij\alpha} B_{\alpha k \beta}
        C_{\beta l \gamma} D_{\gamma mn}.
    \end{align}
    A graphical representation of the triad decomposition can be seen in Fig.~\ref{fig:namedtriads}.
    The tensor network contraction with these can be done efficiently.
    In addition to the two tensor methods, we also use the worm Monte Carlo method~\cite{worm,gattringer}.
    We report results using the original HOTRG, triad HOTRG, and {worm Monte Carlo}.

    \section{Results}
\begin{figure}
    % \begin{minipage}{0.13\textwidth}
        \includegraphics[width=0.49\textwidth]{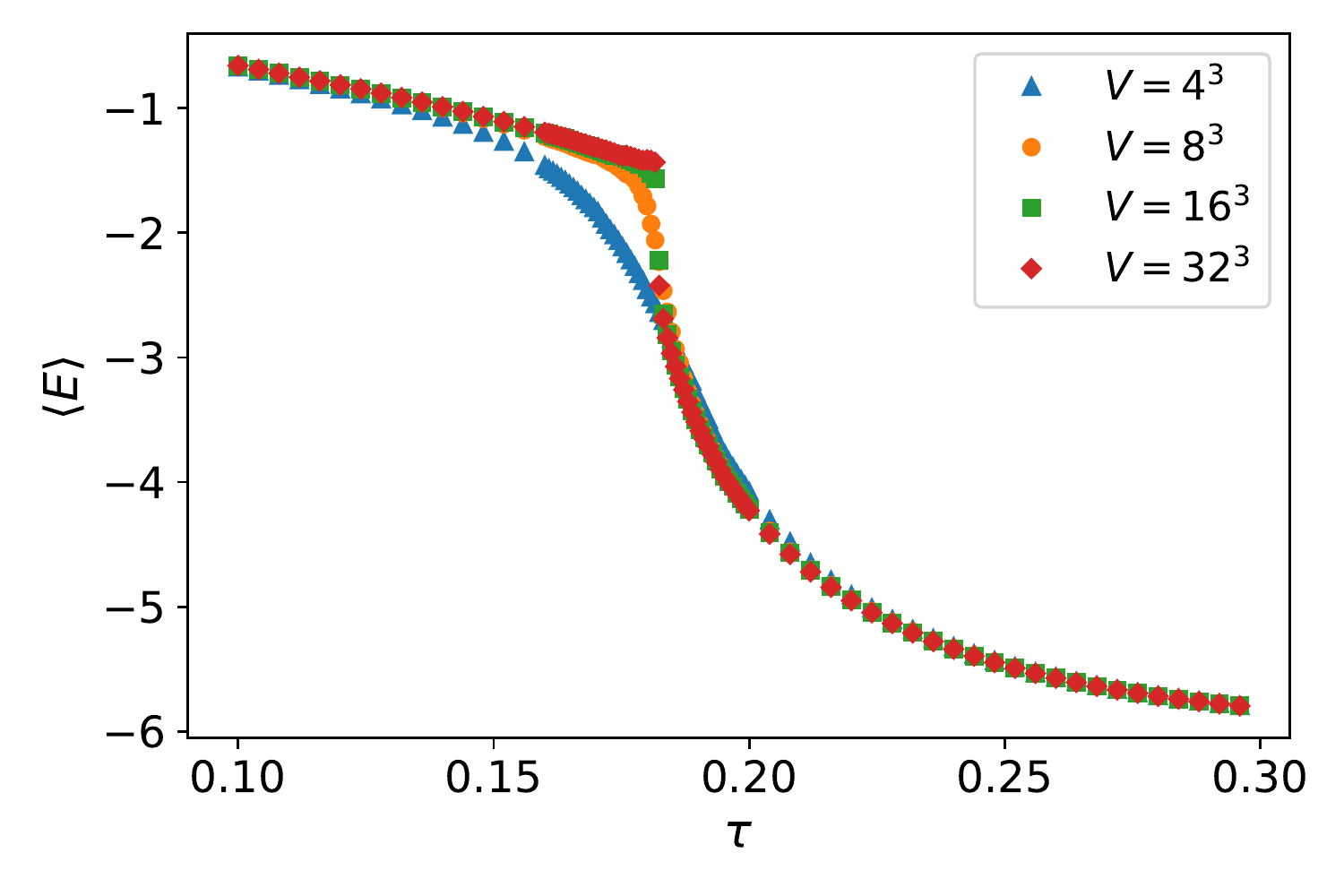}
        % {The average nearest-neighbor Polyakov loop
        % correlator at $\kappa,\mu = 0$.}
        % \end{minipage}\hfill
        % \begin{minipage}{0.13\textwidth}
        \includegraphics[width=0.49\textwidth]{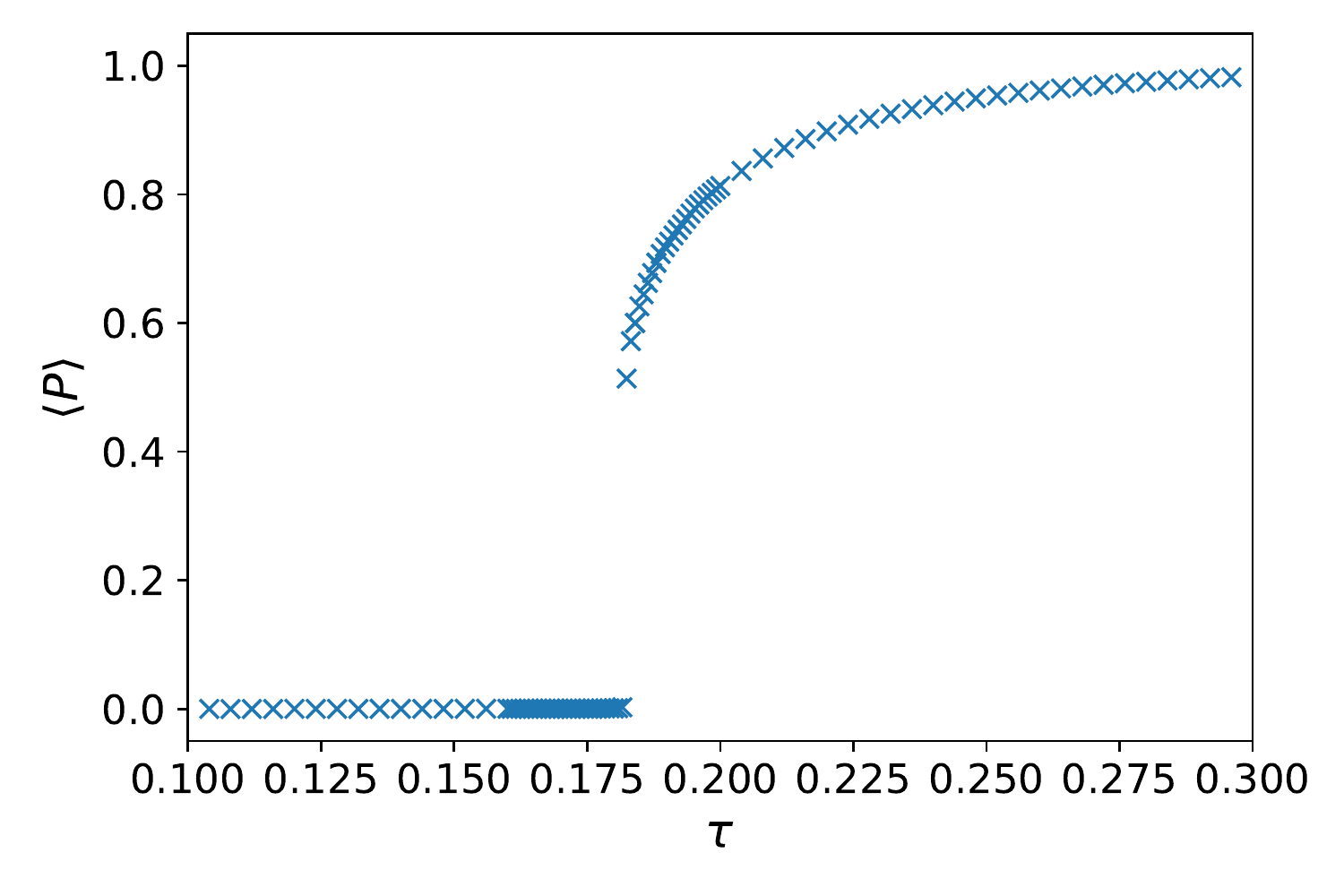}
            % {The average Polyakov loop at $\mu = 0$, $\kappa = 2 \times 10^{-5}$ at $V = 1024^{3}$.}
        % \end{minipage}
        \caption{(left) The average nearest-neighbor Polyakov loop correlator evaluated at $\kappa=0$ as a function of $\tau$ using the TTRG with $D = 40$ states.  Here multiple volumes are shown to demonstrate convergence.  (right) The average Polakov loop as a function of $\tau$, evaluated at $\kappa = 1 \times 10^{-5}$ using the TTRG with $D = 40$ states, at a volume of $V = 1024^{3}$.}
        \label{fig:zerofield}
    \end{figure}

    We calculate the observables mentioned in Sec.~\ref{sec:tensornetwork} using three algorithms: the HOTRG, the TTRG and worm Monte Carlo, and compare them in various circumstances.  In Fig.~\ref{fig:zerofield} we calculate the average nearest-neighbor Polyakov loop correlator at $\kappa = 0$.  We do this using the TTRG with a bond dimension ($D$) equal to 40 for several volumes to demonstrate convergence.  On the right-hand side of Fig.~\ref{fig:zerofield} we show the average Polyakov loop at a small value of $\kappa = 1 \times 10^{-5}$, calculated using the TTRG using $D = 40$.  In both figures we see evidence of a phase transition.
    
\begin{figure}
    % \begin{minipage}{0.13\textwidth}
        \includegraphics[width=0.49\textwidth]{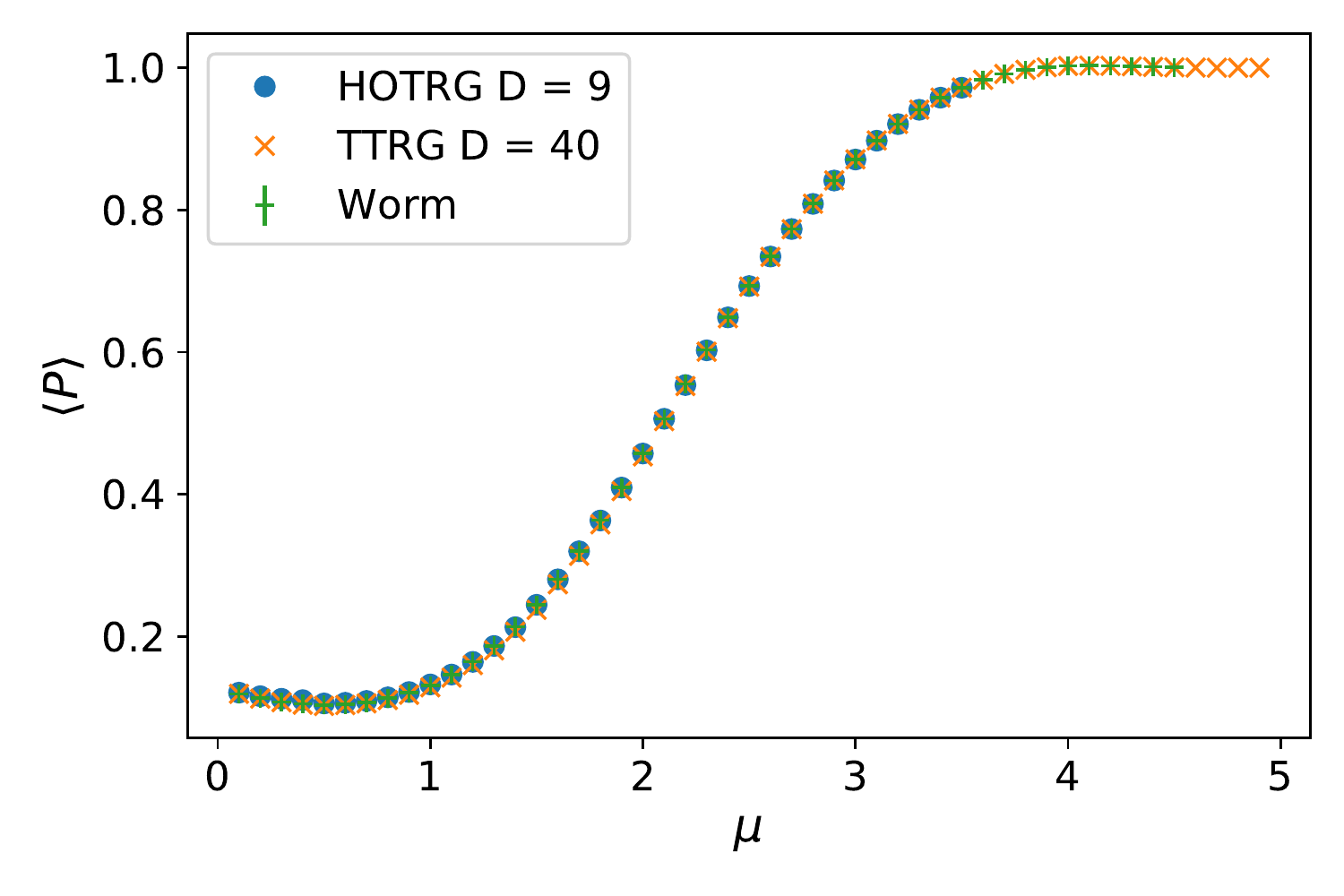}
    %     {The Polyakov loop at $\kappa = 0.05$, and $\tau = 0.1$, $V = 32^{3}$.}
    % \end{minipage}\hfill
    % \begin{minipage}{0.13\textwidth}
        \includegraphics[width=0.49\textwidth]{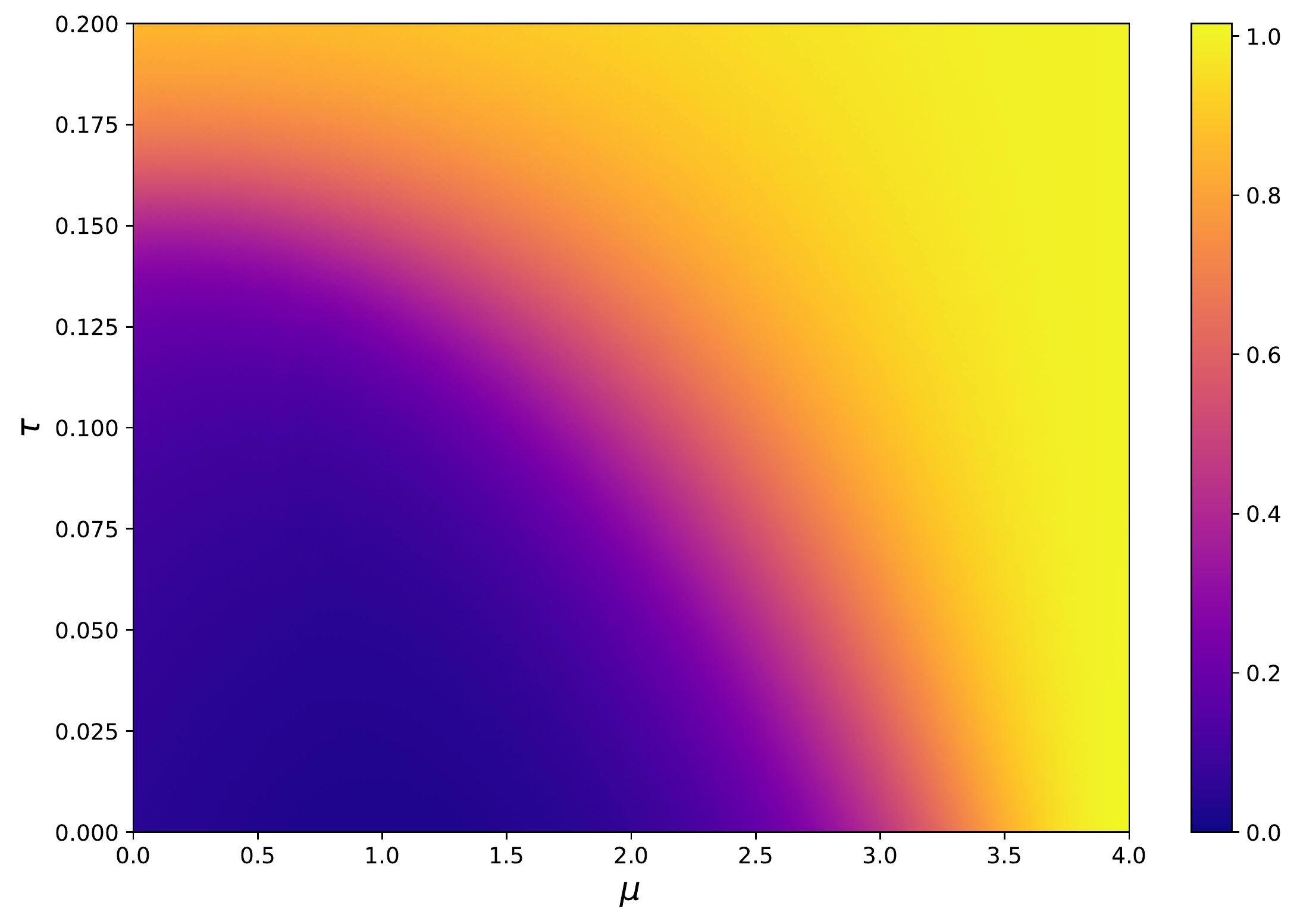}
        % {Heat map of the Polyakov loop at $\kappa = 0.05$  using HOTRG with seven states.}
    % \end{minipage}
        \caption{(left) The average Polyakov loop evaluated at fixed $\tau = 0.1$, and $\kappa = 0.05$ as a function of $\mu$.  Three algorithms are compared: the HOTRG, the TTRG, and the worm Monte Carlo.  We find good agreement between all three methods, and the number of states used in each tensor method is in the legend.  (right) A heat map of the average Polyakov loop in the $\mu$-$\tau$ plane, evaluated at $\kappa = 0.05$, using the HOTRG method with $D = 9$ states.}
        \label{fig:ploop}
    \end{figure}

    In Fig.~\ref{fig:ploop} we calculated the average Polyakov loop at finite $\tau$ and $\kappa$, with $\tau = 0.1$, and $\kappa = 0.05$, for a range of $\mu$ values.  On the left-hand side we see a comparison between the three methods, with all three showing good agreement.  On the right-hand side we plot a heat-map visualization of the average Polyakov loop in the $\tau$-$\mu$ plane, at $\kappa = 0.05$.  This heat-map data was generated using the HOTRG with $D = 9$.

\begin{figure}
    % \begin{minipage}{0.13\textwidth}
        \includegraphics[width=0.49\textwidth]{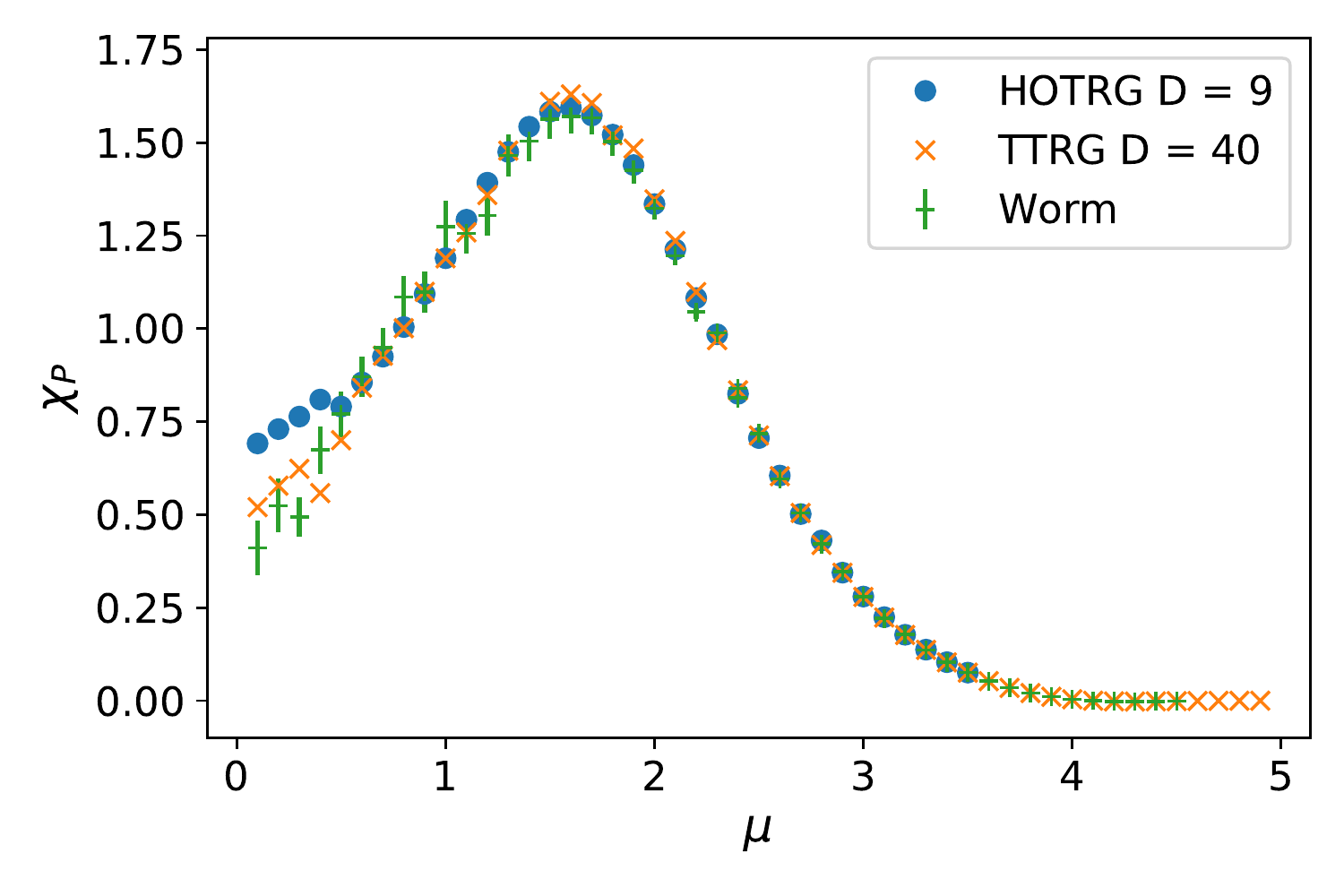}
        % {The Polyakov loop susceptibility at $\kappa = 0.05$, and $\tau = 0.1$, $V = 32^{3}$.}
    % \end{minipage}\hfill
    % \begin{minipage}{0.13\textwidth}
        \includegraphics[width=0.49\textwidth]{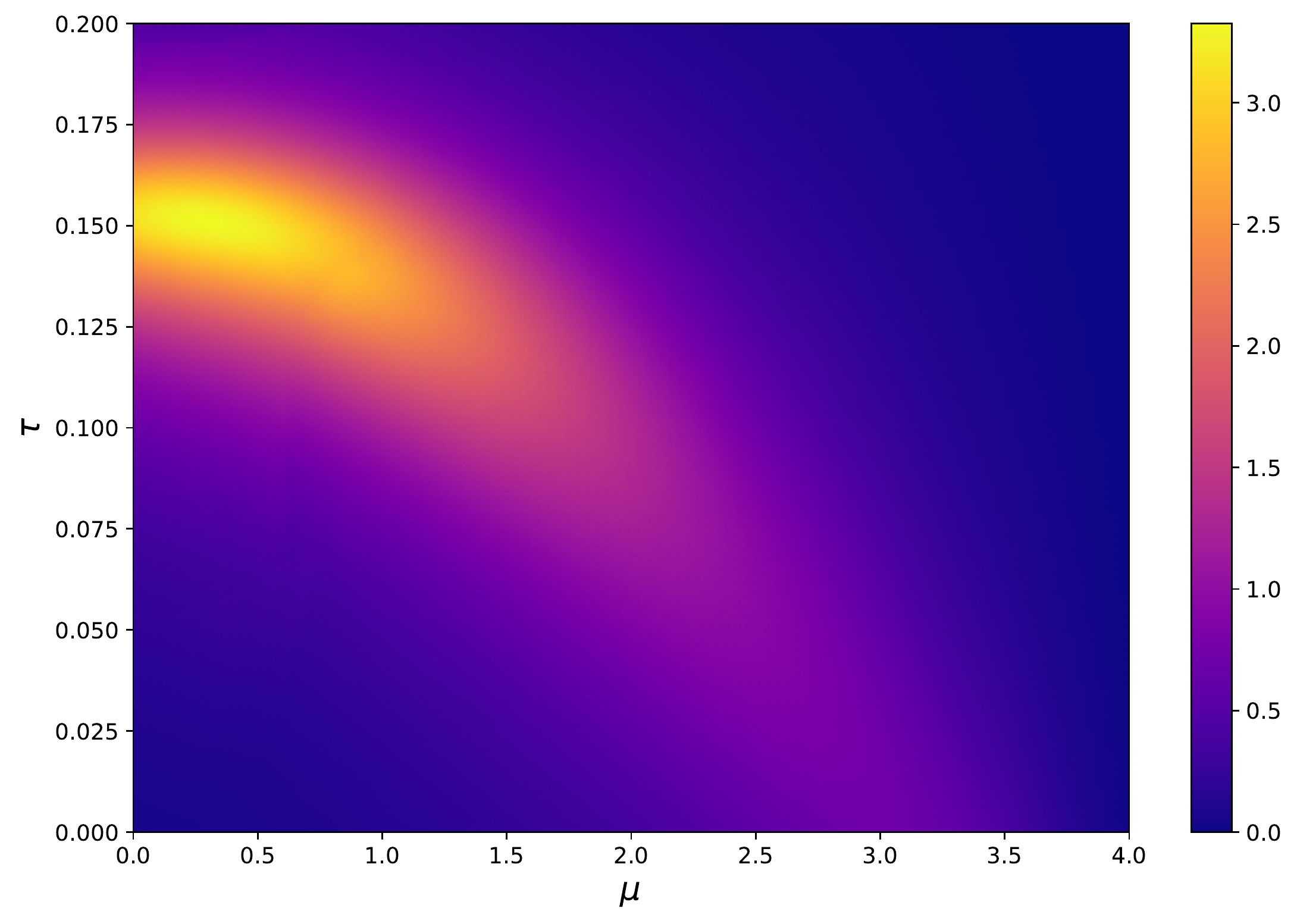}
        % {Heat map of the Polyakov loop susceptibility at $\kappa = 0.05$ using HOTRG with seven states.}
        \caption{(left) The Polyakov loop susceptibility evaluated at $\tau = 0.1$, $\kappa = 0.05$ as a function of $\mu$.  Here three algorithms are compared: the HOTRG, the TTRG, and the worm Monte Carlo.  We find good agreement between the three methods, with noise among all three at small $\mu$.  The number of states for the two tensor methods is found in the legend.  (right) A heat-map of the Polyakov loop susceptibility in the $\mu$-$\tau$ plane evaluate at $\kappa = 0.05$ using the HOTRG method with $D = 9$ states.}
        \label{fig:chi}
    % \end{minipage}
    \end{figure}

    Figure~\ref{fig:chi} shows the Polyakov loop susceptibility.  On the left-hand side we again have a comparison between the three numerical methods, and the three showing good agreement, except at small values of $\mu$ where all three methods are somewhat noisy.  On the right-hand side we plot a heat-map of the susceptibility in the $\tau$-$\mu$ plane at $\kappa = 0.05$.  This data was generated using the HOTRG with $D = 9$.

\section{Conclusion}
We have studied 3+1-dimensional, single-flavor, finite-density QCD in the strong-coupling, high-temperature limit where the model undergoes dimensional reduction to a three-dimensional Abelian spin model.  We studied this effective theory using three different methods: the triad tensor renormalization group, the higher-order tensor renormalization group, and the worm Monte Carlo sampling method.  Where calculations were clean, the three methods agree with each other well.  We found the tensor methods were able to reproduce known features of the phase diagram well; however, calculations at very small values of $\kappa$, and small values of $\mu$ were noisy and unreliable in all three methods.

% \begin{thebibliography}{99}
% \bibitem{hotrg}
% Z. Y. Xie, J. Chen, M. P. Qin, J. W. Zhu, L. P. Yang, and T. Xiang,
% \emph{Coarse-graining renormalization by higher-order singular value decomposition}, \emph{Phys. Rev. B} \textbf{86}, 045139 []

% \bibitem{triad}
% Daisuke Kadoh and Katsumasa Nakayama, (2019),
% `Renormalization group on a triad network',
% \emph{arXiv} hep-lat, 1912.02414.

% \bibitem{...}  

% \end{thebibliography}
\bibliographystyle{JHEP}
%% \bibliography{my-bib-database}

\end{document}